\documentstyle[epsf,art12]{article}
\begin{document}
\title
{Cross sections of $^4$He interaction with protons at 5 GeV/$c$}
\author
{A. V. Blinov, M. V. Chadeyeva, \fbox{V. E. Grechko}, and V. F. Turov}
\date{\it Institute for Theoretical and Experimental Physics (ITEP),
Moscow, Russian Federation}
\maketitle

\begin{abstract}
The total and topological $^4$Hep cross sections as well as
the cross sections of the separate $\alpha$p interaction channels 
and the differential cross sections $\frac{d\sigma}{dt}$ of the 
elastic $\alpha$p scattering were measured using the 2m hydrogen 
bubble chamber exposed to a separated beam of $\alpha$ particles 
from the ITEP synchrotron at 5 GeV/$c$ (the kinetic energy of the 
initial protons in the nuclear rest frame was $T_{\rm p} = 620$~MeV). 
The data obtained have been compared with the results of the previous 
experiments and with the theoretical predictions based on the 
Glauber-Sitenko multiple-scattering theory.
\end{abstract}

The systematic investigations of the nuclear reactions in the 
few-nucleon systems at intermediate energies have been carried out 
in ITEP for the last few years by making use of the lightest nuclei 
($^3$H, $^3$He, $^4$He) as a beam and a liquid-hydrogen bubble 
chambers (80~cm and 2~m) as a target (see e.g. \cite{1} for 
bibliography and background). In \cite{1} we reported the data 
on the cross sections of $^4$He interaction with protons and 
$\alpha$p-elastic scattering below the pion production threshold 
in the elementary NN process. The momentum of the $^4$He nuclei 
averaged over the bubble chamber fiducial volume was equal to 2.7 GeV/$c$ 
(the kinetic energy of the initial protons in the nuclear rest frame 
was $T_{\rm p} = 220$~MeV). Now we present some new data on the $\alpha$p 
interaction cross sections at 5 GeV/$c$ ($T_{\rm p} = 620$~MeV) where 
the pion production is strong enough. The data on the cross sections 
of the separate $^4$Hep interaction channels at this energy range were 
obtained for the first time. The data on the total and differential 
cross sections of the $\alpha$p scattering were compared with the results 
of the previous experiments at the considered energy range as well as 
the Glauber-Sitenko multiple-scattering theory predictions. Note that 
the present experiment is the only one based on non-electronic techniques 
at this energy range.

The 2m liquid-hydrogen bubble chamber was exposed to a separated beam 
of $\alpha$-particles from the ITEP synchrotron at 5~GeV/$c$ momenta. 
About 120000 pictures were received with an average of about 8 initial 
particles for the chamber extension. About 18000 events were measured. 
The peculiarities of the experimental procedure can be found in \cite{1}.

The total cross section is defined as follows
\begin{equation}
\sigma_{tot} = \frac{1}{nl}ln{\frac{1}{1-\frac{N_{int}}{N_0}}}
\end{equation}
where $n$ is the number of hydrogen nuclei in 1~{cm$^3$}, 
$l$ is the fiducial length, $N_0$ is the number of initial tracks, 
$N_{int}$ is the total number of interactions in the fiducial volume 
taking into account the systematic loss of two-prong events.

The two- and three-prong $\alpha$p events were identified with the help 
of kinematic analysis and track ionization measurement. A correction was 
introduced for the loss of two-prong events with a large dip angle. 
This correction was determined from the distribution on the angle 
between the plane of the event and the plane passing through the initial 
track and the vector directed along the magnetic field in the chamber. 
The loss of inelastic two-prong events was $\sim 14\%$. The loss of 
the elastic scattering events with short recoil-proton track for 
$|t| < 0.03$~(GeV/$c$)$^2$, where the differential cross sections 
have not been determined, was evaluated extrapolating the data in the 
interval $0.035 < |t| < 0.1$~(GeV/$c$)$^2$ on the exponential function  
to the region $|t| < 0.03$~(GeV/$c$)$^2$. The differential elastic 
cross section for $|t| < 0.01$~(GeV/$c$)$^2$ has not been measured 
due to poor statistic.

The topological cross sections, the number of events in each channel and 
the cross sections of the $\alpha$p reactions at 5~GeV/$c$ are presented 
in Table 1 (only statistical errors are indicated).

\begin{table}[tbp]
\caption
{Topological cross sections and the cross sections of separate
$\alpha$p interaction channels at 5~GeV/$c$ momentum ($T_{\rm p} = 620$~MeV).} 
\begin{center}
\begin{tabular}{|l|c|l|r|r@{$\pm$}l|}
\hline
Topology & Topological cross & \multicolumn{1}{|c|}{Channel} & Number & 
\multicolumn{2}{|c|}{Cross section} \\
of event & section (mb)  & & of events & \multicolumn{2}{|c|}{(mb)} \\
\hline
 & & $^4$Hep $\rightarrow{} ^4$Hep & 1695 & 31.4 & 2.8 \\
 & & $^4$Hep $\rightarrow{} ^3$Hepn & 2507 & 22.2 & 0.4 \\
 2$^{*}$ & 59.9$\pm$2.8 & $^4$Hep $\rightarrow{} ^4$Hen$\pi^{+}$ & 233 & 2.0 & 0.1 \\
 & & $^4$Hep $\rightarrow{} ^4$Hep$\pi^{0}$ & 295 & 2.6 & 0.2 \\
 & & $^4$Hep $\rightarrow{} ^3$Hepn$\pi^{0}$($\pi^{0}$) & 97 & 0.80 & 0.08 \\
 & & $^4$Hep $\rightarrow{} ^3$Hed$\pi^{0}$ & 59 & 0.50 & 0.07 \\ \hline
 & & $^4$Hep $\rightarrow{} ^3$Hpp & 1952 & 16.1 & 0.4 \\
 & & $^4$Hep $\rightarrow{} $ddp & 335 & 2.8 & 0.2 \\
 & & $^4$Hep $\rightarrow{} $dppn & 2567 & 21.2 & 0.4 \\
 & & $^4$Hep $\rightarrow{} $pppnn($\pi^{0}$) & 1394 & 11.5 & 0.3 \\
 & & $^4$Hep $\rightarrow{} ^3$Hd$\pi^{+}$ & 101 & 0.80 & 0.08 \\
3$^{*}$ & 60.2$\pm$0.7 & $^4$Hep $\rightarrow{} ^3$Hpn$\pi^{+}$ & 362 & 3.0 & 0.2 \\
 & & $^4$Hep $\rightarrow{} $ddn$\pi^{+}$ & 79 & 0.65 & 0.07 \\
 & & $^4$Hep $\rightarrow{} $dpnn$\pi^{+}$($\pi^{0}$) & 128 & 1.10 & 0.09 \\
 & & $^4$Hep $\rightarrow{} $ppnnn$\pi^{+}$($\pi^{0}$) & 79 & 0.65 & 0.07 \\
 & & $^4$Hep $\rightarrow{} ^3$Hpp$\pi^{0}$ & 117 & 1.00 & 0.09 \\
 & & $^4$Hep $\rightarrow{} $ddp$\pi^{0}$ & 53 & 0.40 & 0.06 \\
 & & $^4$Hep $\rightarrow{} $dppn$\pi^{0}$($\pi^{0}$) & 93 & 0.80 & 0.08 \\ \hline
4$^{*}$ - 5$^{*}$ & 1.4 $\pm$ 0.1 &  &  &\multicolumn{2}{|c|}{ } \\ \hline
\end{tabular}
\end{center}
\end{table}

The total $\alpha$p-interaction cross section is equal to 121.5$\pm$2.9~mb 
(the error is statistical only). The systematic error in the absolute 
normalization of the cross section is $\sim$ 3\%.

The differential cross sections of the $\alpha$p elastic scattering at 5~GeV/$c$ 
are presented in Table 2. These data can be very well parameterized by the 
exponential function $\frac{d\sigma}{dt} = Ae^{Bt}$ 
with $A = (9.4 \pm 0.5)\times 10^2$~mb/(GeV/$c$)$^2$ and 
$B = 31 \pm 1$~(GeV/$c$)$^{-2}$ ($\chi ^2/NF = 0.4$).

\begin{table}[tbp]
\caption
{Differential cross sections
of the $\alpha$p elastic scattering
at 5~GeV/$c$ ($T_{\rm p} = 620$~MeV).} 
\begin{center}
\begin{tabular}{|c|r@{$\pm$}l|}
\hline
  -t, (GeV/$c$)$^2$ & \multicolumn{2}{|c|}{$\frac{d\sigma}{dt}$,
 mb/(GeV/$c$)$^2$} \\ \hline
 0.035 & 311 & 19 \\
 0.045 & 246 & 15 \\
 0.055 & 172 & 13 \\
 0.065 & 134 & 12 \\
 0.075 & 97 & 10 \\
 0.085 & 70 & 8 \\
 0.100 & 52 & 5 \\ \hline
\end{tabular}
\end{center}
\end{table}

For the theoretical interpretation of the elastic scattering data we use 
the Glauber-Sitenko multiple-scattering theory taking into account the 
spin-isospin structure of the NN scattering amplitude which is the simple 
generalization of the conventional Glauber theory for non-diffractive NN scattering. 

The details of the present theoretical approach are given in \cite{1}. 
In our present calculations we also use a parameterization of the $^4$He 
nuclear ground-state density in the gaussian form $\rho(r) \sim exp(-r^2/R^2)$ 
with $R = 1.25$~fm \cite{1}.

Figure~\ref{f1} shows the differential cross sections 
$\frac{d\sigma}{dt}$ of the
$^4$Hep elastic scattering at 5 GeV/$c$ momentum ($T_{\rm p} = 620$~MeV) 
and the results of calculations in the framework of the Glauber-Sitenko 
multiple-scattering theory taking into account the spin-isospin structure 
of the NN scattering amplitude (curve). For comparison, the p$^4$He
elastic scattering data at $T_{\rm p} = 500$~MeV 
\cite{2}, $560$~MeV \cite{3}, $587$~MeV \cite{4}, $600$~MeV \cite{5}, 
and $695$~MeV \cite{6} are also shown here. As can be seen from the Figure 
the shapes of the experimental distributions for all these experiments are 
quite similar. All data are in good agreement. Note that the independent 
experiments based on the different experimental techniques give  very close 
results. The multiple scattering theory perfectly describes the shape of the 
distribution but the theoretical curve is slightly below the experimental data.

The total $\alpha$p cross section obtained in this experiment at 
$T_{\rm p} = 620$~MeV ($\sigma_{tot} = 121.5 \pm 2.9$~mb) is in good agreement 
with the result of \cite{7} at $T_{\rm p} = 563$~MeV, 
$\sigma_{tot} = 123.7 \pm 0.9$~mb. Some discrepancy with the theoretical 
value for total cross section calculated through the optical theorem 
($\sigma_{tot}^{th} = 113.4$~mb at $T_{\rm p} = 620$~MeV) is caused by 
the accuracy of the theoretical scheme used here (see \cite{1} for details).

The main results of the paper are as follows.

(i) Using a 2m ITEP liquid-hydrogen bubble chamber exposed to a separated 
beam of $\alpha$ particles with 5~GeV/$c$ momentum ($T_{\rm p} = 620$~MeV) 
we have measured the total, topological $\alpha$p cross sections as well as 
the cross sections of separate $\alpha$p interaction channels. The data on 
the cross sections of the $\alpha$p reaction channels at this energy range 
were obtained for the first time. The value of total $\alpha$p cross section 
obtained in present experiment is in good agreement with the data from the literature.

(ii) The data on the $\alpha$p differential cross sections are in good 
agreement with the previous data at the considered energy range and with 
the  multiple-scattering theory predictions as well.

The authors thank the stuff of the 2m hydrogen bubble chamber, the laboratory 
of the separated beams, the laboratory of the cryogenic engineering that took 
part in the experiment as well as the personnel of the scanning devices and 
computers. We are very grateful to I. V. Kirpichnikov for his permanent 
assistance throughout this work.

\begin{figure}
\vskip -2cm
\epsfysize=20cm
\centerline{\epsffile{Elastf.eps}}
\vskip -6cm
\caption{The differential cross section of the reaction $^4$Hep
$\rightarrow$ $^4$Hep.  $\bullet$ --- our data ($T_{\rm p} = 
620$~MeV);
$\Diamond$ --- data from ref.~\protect\cite{2} ($T_{\rm p} = 500$~MeV);
$\bigtriangleup$ --- data from ref.~\protect\cite{3} ($T_{\rm p} = 
560$~MeV); $\bigtriangledown$ --- data from ref.~\protect\cite{4} 
($T_{\rm p} = 587$~MeV); $\Box$ --- data from ref.~\protect\cite{5} 
($T_{\rm p} = 600$~
MeV); $\circ$ --- data from ref.~\protect\cite{6} ($T_{\rm p} = 
695$~MeV).  The solid line corresponds to the theoretical predictions 
(at $T_{\rm
p} = 620$~MeV) based on the Glauber-Sitenko multiple-scattering theory
taking into account the spin-isospin structure of the NN scattering
amplitude.} 
\label{f1}
\end{figure}

\end{document}